# Anomalous refraction based on acoustic metasurfaces with membranes

Shilong Zhai, Changlin Ding, Huaijun Chen, Song Liu, Chunrong Luo and Xiaopeng Zhao

*Smart Materials Laboratory, Department of Applied Physics, Northwestern Polytechnical University, Xi'an 710129 P. R. China*

E-mail: xpzhao@nwpu.edu.cn

**Abstract**

The investigation of metasurface, which is of great current interest, has opened up new degrees of freedom to research metamaterials. In this paper, we propose an ultrathin acoustic metasurface consisting of a series of structurally simple drum-like subunit cells. Eight units with different sizes are selected to realise discrete transmission phase shifts ranging from 0 to $2\pi$ at 3.5 KHz. The designed metasurface is capable of manipulating sound waves at will, including anomalous refraction and conversion of propagating waves into surface waves, which can be predicted precisely by the generalized Snell's law. It is worth noting that the total thickness of such planar metasurface is approximately $\lambda/5$, which may be beneficial to the miniaturization and integration of acoustic equipment.

## 1. Introduction

Metamaterials as artificial structures possess many unique properties compared to conventional materials as they can control the propagation of waves arbitrarily [1-12]. Among numerous types of metamaterials, there is one species of structure, referred to as metasurface, is able to modulate the orientation of wavefront with a very simple mechanism. Besides, the thickness of this model is generally deep-subwavelength. Recently, metasurfaces in electromagnetics have given rise to various novel phenomena, for an example, planar lens for manipulating wavefront [13, 14]. Since electromagnetic waves and sound waves share many important wave characteristics, it is expected that acoustic metasurfaces, being capable of controlling the direction of sound waves, can also be constructed.

To date, many different approaches have been proposed to achieving acoustic metasurfaces. For instance, Zhao *et al* [15] accomplished the redirection of reflected waves via manipulating the impedance distribution of a flat metasurface reflector. To overwhelm the challenge of impedance-mismatch, Li *et al* [16] designed a hybrid labyrinthine structure by modifying refractive indexes at the interface, and realised the convergence of sound waves in three-dimensional space utilizing a planar acoustic lens. Another kind of method to tailor acoustic wavefronts is to modulate the phase of waves in the transverse direction of metasurfaces. Li and Tang *et al* [17, 18] obtained anomalous acoustic reflection and extraordinary refraction phenomena through coiling-up-space structures. As for the effects of negative refraction, there are also many other examples taking advantage of metamaterials [19-21].

Despite the described advantages, aforementioned acoustic metasurfaces inevitably suffer from the restriction of fabrication due to the complication of inner microstructures of coiling units. Moreover, the impedance of acoustic rigid wall used for increasing propagation path of sound waves mismatches with that of surrounding air, which unavoidably enlarges dissipation in metasurfaces.

The elastic membrane with fixed boundaries, as a type of acoustic metamaterial, exhibits various unique properties owing to the membrane's weak elastic moduli [22-24]. Whereas, most people care about the huge transmission loss of membranes within the frequency range between two eigenmodes. In a recent report, it has been demonstrated that a thin membrane metamaterial can achieve almost total transmission of sound waves in air [25]. On the basis of this feature, in this paper, membranes are introduced into the construction of acoustic metasurfaces in an audible regime. A drum-like subunit, containing a cavity filled with air column and two membranes whose boundaries are fixed at the ends of cavity, is proposed. Acoustic metasurfaces constructed with such simple subunits exhibit high-efficiency performances in steering acoustic wavefronts. Numerical simulations demonstrate that the transmission phase of this subunit can be modulated from 0 to $2\pi$ at will by appropriately optimizing its structure size. Consequently, the orientation of refracted waves can be manipulated via spatially varying the dimensions of subunits. As the incident angle of sound wave changes from -90° to 90°, extraordinary phenomena including the abnormal refraction and conversion of propagating waves into surface waves are obtained in comparison with conventional materials. In addition, the thickness of considered metasurface is only about $\lambda/5$, which is promising for the miniaturization and integration of acoustic devices.

**2. Description of model**

Figure 1(*a*) displays the schematic diagram of the acoustic metasurface, which is composed of a cavity filled with air and two membranes with boundaries fixed on the ends of side walls to seal the cavity. It can be seen that the present structure is identical with a drum. Acoustic beams impinge upon the front face of metasurface (dark green arrows), and emit from the rear side after passing through the air medium sealed in the cavity. The partial amplification drawing of membrane is given in the inset of figure 1(*a*). The mass density, Yang's modulus, and Poisson's ratio for membrane are 920 kg/m$^3$, 6.9×10$^9$ Pa, and 0.36,

respectively. The width *w* of membrane is an adjustable parameter, and the edge of membrane is fixed to the side wall. Membranes will not deform when there is no sound wave impinging onto it, but will vibrate when driven by the sound pressure of incident waves. A sketch of forced deformation is illustrated by the iridescent curve in figure 1(a). The dark green and fluorescent green arrows refer to the propagation orientation of incident and transmitted waves, respectively. Intuitively, the proposed structure will introduce extra propagating phase delay of acoustic waves than air, which will be demonstrated by the following simulated results.

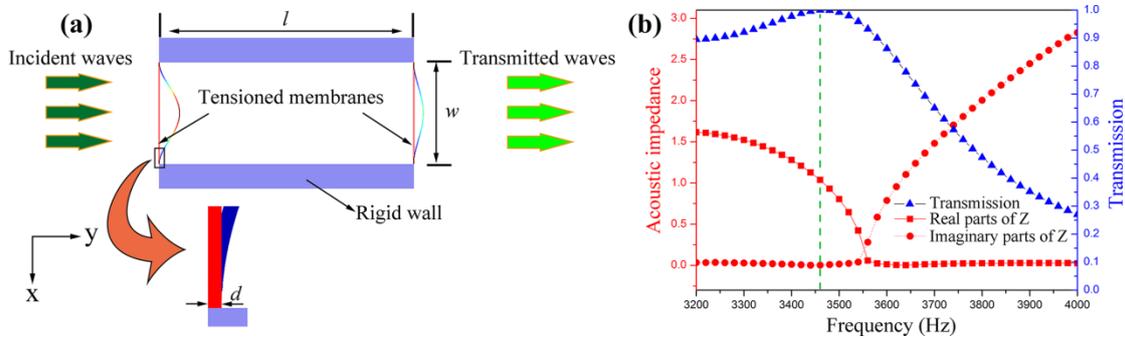

Figure 1. (a) Schematic of the subunit of proposed metasurface. The purple rectangles depict rigid walls with length *l* =20mm. The red lines represent tensioned membranes with thickness *d* = 0.07 mm (clearly displayed in the inset). (b) The transmission (triangular dots), and the real part (square dots) and imaginary part (circular dots) of effective acoustic impedance *Z* of subunit (with membrane width *w* = 8.271 mm) as a function of frequency.

For the purpose of designing an acoustic metasurface with high transmission efficiency, it is crucial to keep the acoustic impedance of subunit matching with that of air. It will be demonstrated in the following portion that the impedance of reported drum-like subunit is approximate to that of background in a certain frequency range by appropriately adjusting the size of microstructure (i.e., the width *w* of membrane). This microstructure, at low frequencies, can be regarded as a homogenized medium. The effective acoustic impedance of metasurface can be derived from transmission and reflection results by retrieving the effective parameter from a homogeneous medium [26]. Figure 1(*b*) indicates the numerically calculated acoustic impedance $Z$ of subunit with $w$ = 8.271 mm as a function of frequency. The red lines with square dots and circular dots are real parts and imaginary parts of $Z$, respectively. It is observed that good impedance matching is achieved near 3.46 KHz ($Z$ = 1). The power transmission coefficient also supports this phenomenon, which shows almost total transmission near 3.46 KHz, as illustrated in figure 1(*b*). It is also found that for the frequencies above 3.55 KHz, the real parts of $Z$ are close to 0, while the imaginary parts increase gradually, which implies the augment of attenuation and decline of transmission. In this study, we choose 3.5 KHz as the working frequency.

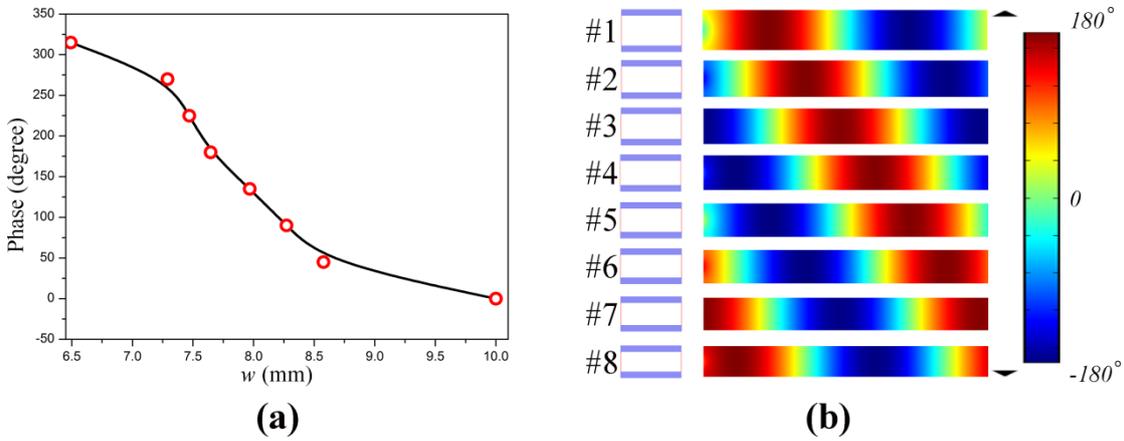

**(a)** **(b)**

Figure2. (a) The phase of transmitted wave as a function of the width $w$ of membrane with the chosen wavelength of incident wave $\lambda$ = 98 mm. The red circles refer to discrete phases of eight subunits having specific membrane widths. The phase shift between two adjacent subunits is kept to be 45°. (b) The left portion denotes the sketch of eight subunits, and the right portion illustrates the corresponding field pattern of transmitted phase. The phase changes can be observed directly.

In the following demonstrations, we will show that the generalized Snell's law can be realised through the present metasurface. By appropriately selecting the geometrical sizes, eight subunits are able to provide discrete phase changes ranging from 0 to $2\pi$ with an interval of $\pi/4$ between adjacent subunits. The widths of membranes for eight subunits are optimized to be 10 mm, 8.577 mm, 8.271 mm, 7.969 mm, 7.645 mm, 7.469 mm, 7.29 mm, and 6.492 mm, respectively. The phase for transmitted waves as a function of membrane width is plotted in figure 2(a). To interpret the gradient variation of phase more directly, figure 2(b) shows the phase field distributions of transmission for eight subunits. It can be clearly observed that these eight subunits are sufficient for realising phase shifts covering $2\pi$. It is worth pointing that the proposed metasurface is ultrathin with thickness approximately equal to $\lambda/5$. In what follows, we will verify that the constructed metasurface can manipulate the direction of transmitted acoustic beams.

## 3. Design of different phase gradients for abnormal refraction

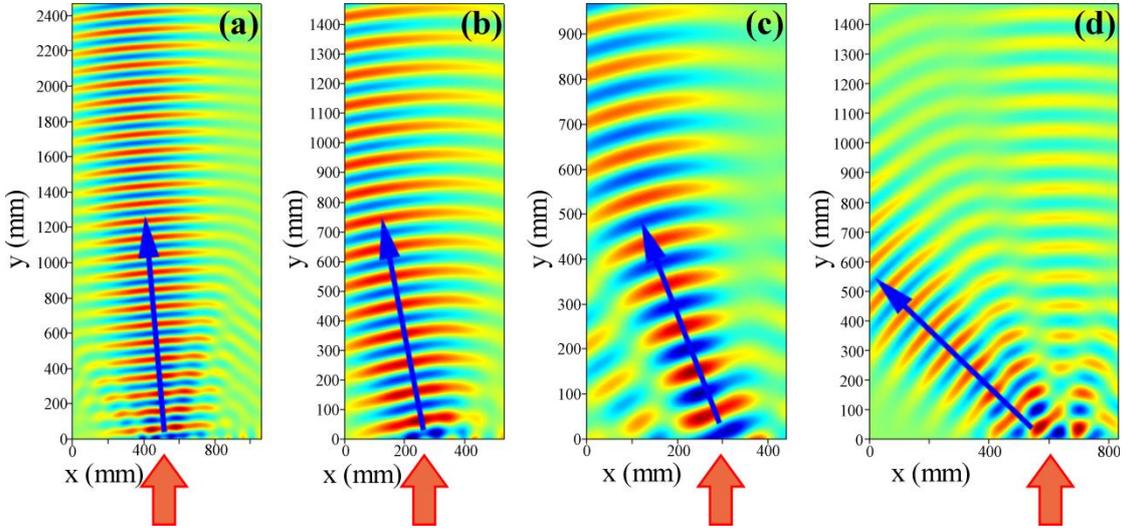

Figure 3. The simulated snapshots of pressure map for refracted beams (incident waves propagate along the direction perpendicular to the interfaces). Phase gradients $d\varphi/dx$ for (a), (b), (c), and (d) are 5.582 rad/m, 11.122 rad/m, 24.511 rad/m, and 45.295 rad/m, respectively. The orange arrows refer to the direction of incident waves with a Gaussian beam. It can be seen that the refracted angle changes with phase gradient.

The refracted angles are certainly determined by the classical Snell's law, due to the conservation of momentum along the tangential orientation of interface. However, if an abrupt phase change is artificially introduced on the surface, an extra term depicting the phase gradient ($d\varphi/dx$) should be added to generalize the Snell's law:

$$\frac{2\pi}{\lambda}\sin\theta_{out} = \frac{2\pi}{\lambda}\sin\theta_{in} + \frac{d\varphi}{dx}, \qquad (1)$$

where $\theta_{in}$ and $\theta_{out}$ represent the incident and refracted angle, respectively. Equation (1) indicates that the refracted angle can be modulated at will by selecting appropriate phase profiles along surface. The $\theta_{out}$ can be deduced from equation (1) as follows:

$$\theta_{out} = acr\sin(\sin\theta_{in} + \frac{\lambda}{2\pi}\frac{d\varphi}{dx}). \qquad (2)$$

To verify this feature, four phase gradient profiles (5.582 rad/m, 11.122 rad/m, 24.511 rad/m, and 45.295 rad/m) are chosen under the condition of normal incidence $\theta_{in} = 0°$. The theoretical refracted angles derived from equation (2) are 5°, 10°, 22.5°, and 45°, respectively. Simulations of the metasurfaces mentioned above are performed with the COMSOL Multiphysics. Figures 3(*a*)-(*d*) display the snapshots of pressure map of transmitted beams for these four phase configurations. A sound source is set to generate the plane wave with a Gaussian distribution. Refracted angles can be obtained from the wavefront direction of transmitted beams. In figure 4, it is of significance that the simulated results (red square dots) are in good qualitative agreement with theoretical values (black line) calculated via equation (2).

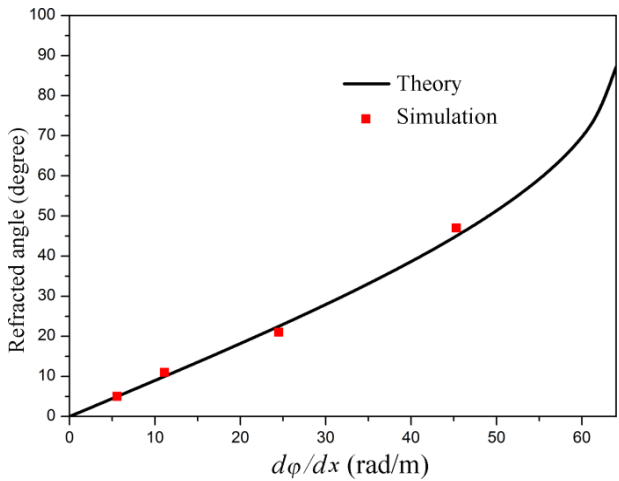

Figure 4. The refracted angle as a function of phase gradient profile (the incident angle is kept to be 0°). The black line indicates calculated results from the generalized Snell's law, and the red square dots denote the results obtained from figure 3.

The black line in figure 5 refers to the energy conversion efficiency for refracted beam with phase profile $d\varphi/dx$ = 5.582 rad/m, defined as a ratio of transverse integration of $|p|^2$ of transmitted beam (along the line 1000 mm away from the back side of metamaterial) to that of incident beam. Although this configuration is only designed for the frequency of 3.5 KHz, the efficiency is over 80% between 3.45 KHz and 3.7 KHz. The high conversion can be roughly explained as follows: in this frequency range, the phase gradients of adjacent subunits are no longer equal ($\pi/4$), nevertheless, there are still phase shifts among them, which consequently leads to positive transversal momentums. This phenomenon can be attributed to the average effect of all subunits for the broadening of operating frequency. It can be verified from the insets in figure 5 that abnormal refractions still exist at different frequencies, whereas the refracted angles deviate slightly from that at 3.5 KHz.

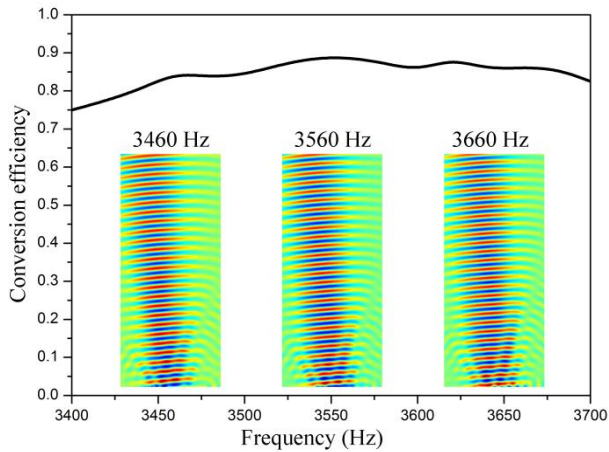

Figure 5. The energy conversion efficiency as a function of frequency when $d\varphi/dx$ is fixed to be 5.582 rad/m with a Gaussian beam impinging normally upon the surface. Insets: the temporal pressure fields of transmitted waves at three different frequencies.

## 4. Relation between refracted angles and incident angles

When the phase configuration is fixed to be $d\varphi/dx$ = 45.295 rad/m, the refracted angle can be calculated accurately from incident angle according to the generalized Snell's law. For example, the refracted angle $\theta_{out}$ = 45° corresponds to the incident angle $\theta_{in}$ = 0°, as manifested in figure 6(*a*). In addition, the $\theta_{out}$ increases with the increment of $\theta_{in}$. However, as $\theta_{in}$ reaches a certain critical angle $\theta_{crit}$, $\theta_{out}$ will increase to 90°, implying that the propagating wave is converted into surface wave whose surface wave number $k_x$ is exactly the same as $k_0$. At this moment, the propagating wave vector along Y-axis is $k_y = \sqrt{k_0 - k_x} = 0$. In this situation, the critical incident angle $\theta_{crit} = arc\ \sin(1-45.295/k_0)/6.28 \times 360° = 17.06°$. In other words, the refracted angle increases monotonically when incident angle increases from -90° to 17.06°, as illustrated in figure 6(b) (the left side of gray region). The red square dots indicate simulated results extracted from the polar plots of refracted beams, as shown in figure 6(a), and the black solid line represents theoretical curves calculated by equation (2). It is seen that simulated results follow almost exactly the theoretical predictions. Note that, along with the variation of $\theta_{in}$ from -45° to 0°, the sign of refracted angle is contrary to that of incident angle, which means both of them are located on the same side of normal. Figure 8(a) describes the simulated transient image of sound pressure for incident angle of -10° and exhibits the negative refraction phenomenon clearly. For the incident angles ranging from 0° to 17.06°, $\theta_{out}$ is on the opposite side of normal with $\theta_{in}$, as manifested in figure 8(b).

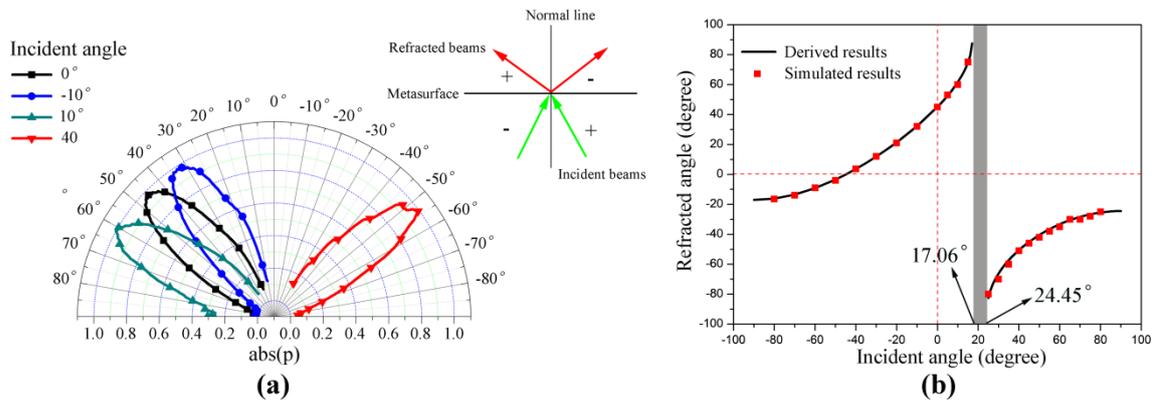

Figure 6. (a) A polar plot of simulated acoustic pressure fields of transmitted beams (1000 mm away from metasurface) at different incident angles. The phase gradient is chosen to be 45.295 rad/m. A Gaussian beam is incident onto the surface. The amplitudes of acoustic pressures are normalized. It is regulated in this paper that, if the incident beams are on the right side of normal, the sign of $\theta_{in}$ is '+'; otherwise the sign is '-'. Rules for $\theta_{out}$ are as opposed to that for $\theta_{in}$, as shown in the inset. (b) The variation of $\theta_{out}$ with $\theta_{in}$. The black line and red square dots refer to theoretical and simulated results, respectively. The gray region indicates that the refracted angles do not exist (i.e., the metasurface is opaque for incident angles ranging from 17.06° to 24.45°).

Further increasing $\theta_{in}$ will result in surface wave, of which the wave number $k_x$ will be larger than $k_0$. In such a case, the wave vector component $k_y$ is imaginary, i. e. the sound waves propagating along Y-axis are evanescent waves. For instance, when the incident angle is 20°, the surface wave number $k_x = k_0 \sin\theta_{in} + d\varphi/dx = 67.2$ rad/m, which is larger than that in air ($k_0 = 64.08$ rad/m). As a result, the transmitted waves can only travel along the surface. Figure 7 displays a pressure field pattern of transmitted waves as the angle of incident beam is 20°. It is found that the propagating waves are converted into surface waves along X-axis (red arrow) after incident waves impinging upon the sample (blue arrows).

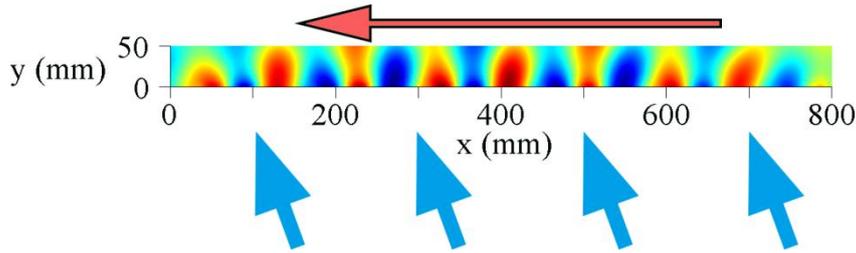

Figure 7. Conversion of propagating waves into surface waves near the metasurface with phase gradient $d\varphi/dx = 45.295$ rad/m. The blue and red arrows indicate the direction of incident and refracted beams respectively with incident angle $\theta_{in} = 20°$.

It is worth to note that, as the phase profile $d\varphi/dx = 45.295$ rad/m, the period of metasurface (138.648 mm) is larger than the working wavelength (98 mm). In this situation, the non-local effects generated by periodicity cannot be neglected. Thus the proposed metasurface can also be regarded as a 1D phononic crystal that is able to change the phase of sound waves. The lattice element is constructed of proposed subunits with period $T = 138.648$ mm, and the created phase gradient is $d\varphi'/dx = 2\pi/T$. Under this assumption, the generalized Snell's law can be modified as follows:

$$\frac{2\pi}{\lambda}\sin\theta_{out} = \frac{2\pi}{\lambda}\sin\theta_{in} + \frac{d\varphi}{dx} + m\frac{d\varphi'}{dx}, \qquad (3)$$

where $m$ is determined from the momentum matching on surface. For incident angles below $\theta_{crit}$, $m = 0$. When $\theta_{in}$ is above $\theta_{crit}$, the equation (3) can be satisfactorily fitted with $m = -3$. In this case, there is another critical angle $\theta'_{crit} = 24.45°$. As $\theta_{in}$ increases from $\theta_{crit}$ to $\theta'_{crit}$, the refracted angles do not exist (as depicted by the gray region in figure 6(b)), which implies that the transmitted acoustic beams are surface waves. For the case of $\theta_{in} > \theta'_{crit}$, negative refraction effects appear again, as shown in figures 6(b) and 8(c).

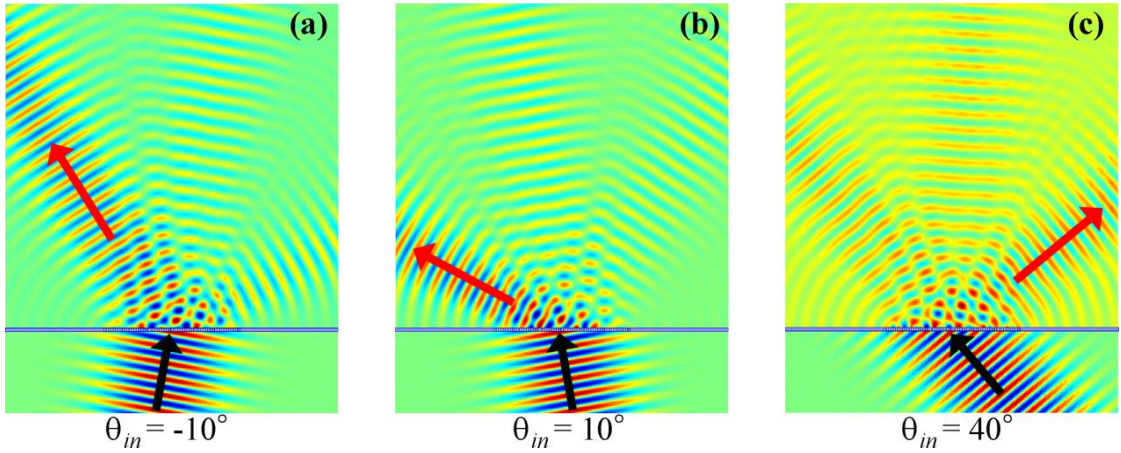

Figure 8. Snapshots of pressure field for three different incident angles (-10°, 10°, and 40°) with phase profile $d\varphi/dx$ = 45.295 rad/m. The black arrows represent the plane waves incident with Gaussian beams, and the red arrows denote transmitted beams. It can be seen that, apart from the transmitted waves propagating along red arrows, there are also a few parts of energy traveling along other directions, which has been interpreted in detail in reference [27].

## 5. Conclusion

In summary, we numerically demonstrate a novel acoustic metasurface composed of simple subunits to manipulate refracted beams at will. In order to realize the transmission phase shifts from 0 to $2\pi$, the subunits are made up of two-layer elastic membranes and a cavity with different dimensions. The matching of acoustic impedance of metasurfaces with that of air allows high transmission efficiency. With the appropriate modulation of phase gradients and change of incident angles, extraordinary properties such as abnormal refractions and conversion of propagating waves into surface waves can be accomplished. With the simplicity of constructing these metasurfaces, a similar design can be expanded to three-dimensional case. The strategy presented in this work provides a new clue for the preparation of ultrathin slab lenses and acoustic switches.

**Method of simulations**

All simulated results in this paper are obtained via the finite Element Method based on commercial software COMSOL Multiphysics$^{TM}$ 4.3 b. Materials used in the simulation are air as a propagation medium, side walls as acoustic rigid boundaries, and elastic membranes. Plane wave radiation and the periodic boundary condition are imposed for the calculation of transmission phases of subunits. For the simulation of transient pressure fields, background pressure field with a Gaussian shape is employed, and perfectly matched layers are utilized to eliminate the interference from reflected waves.

**Acknowledgments**

This work was supported by the National Natural Science Foundation of China under Grant Nos. 50936002, 11174234 and 51272215 and the National Key Scientific Program of China (under project No. 2012CB921503)

## Figure captions

Figure 1. (a) Schematic of the proposed subunit of metasurface. The purple rectangles depict rigid walls with length $l$ =20mm. The red lines represent tensioned membranes with thickness $d = 0.07$ mm (clearly displayed in the inset). (b) The transmission (triangular dots), and the real part (square dots) and imaginary part (circular dots) of effective acoustic impedance $Z$ of subunit (with membrane width $w = 8.271$ mm) as a function of frequency.

Figure2. (a) The phase of transmitted waves as a function of the width of membrane $w$ with the chosen wavelength of incident wave $\lambda = 98$ mm. The red circles refer to discrete phases of eight subunits having specific membrane width. The phase shift between two adjacent subunits is kept to be 45°. (b) The left portion denotes sketches of eight subunits, and the right portion illustrates the corresponding phase field pattern of transmitted waves. The phase changes can be observed directly.

Figure 3. The simulated snapshots of pressure map for refracted beams (incident waves propagate along the direction perpendicular to interfaces). Phase gradients $d\varphi/dx$ for (a), (b), (c), and (d) are 5.582 (rad/m), 11.122 (rad/m), 24.511 (rad/m), and 45.295 (rad/m), respectively. The orange arrows refer to the direction of incident waves with a Gaussian beam. It can be seen that the refracted angles change with phase gradients.

Figure 4. The refracted angle as a function of phase gradient profile (the incident angle is kept to be 0°). The black line indicates calculated results from the generalized Snell's law, and the red square dots denote the results

from figure 3.

Figure 5. The energy conversion efficiency as a function of frequency when $d\varphi/dx$ is fixed to be 5.582 (rad/m) with a Gaussian beam impinging normally upon the surface. Insets: the temporal pressure fields of transmitted waves at three different frequencies.

Figure 6. (a) A polar plot of simulated acoustic pressure fields of transmitted beams (1000 mm away from metasurface) at different incident angles. The phase gradient is chosen to be 45.295 (rad/m). A Gaussian beam is incident onto the surface. The amplitudes of acoustic pressure are normalized. It is regulated in this paper that, if the incident beams are on the right side of normal line, the sign of $\theta_{in}$ is '+'; otherwise the sign is '-'. Rules for $\theta_{out}$ is as opposed to that for $\theta_{in}$, as shown in the inset. (b) The variation of $\theta_{out}$ with $\theta_{in}$. The black line and red square dots refer to theoretical and simulated results respectively. The gray region indicates that the refracted angles do not exist (i.e., the metasurface is opaque for incident angles ranging from 17.06° to 24.45°).

Figure 7. Conversion of propagating waves into surface waves near the metasurface with phase gradient $d\varphi/dx$ = 45.295 (rad/m). The blue and red arrows indicate the direction of incident and refracted beams respectively with incident angle $\theta_{in}$ = 20°.

Figure 8. Snapshots of pressure field for three different incident angles (-10°, 10°, and 40°) with phase profile $d\varphi/dx$ = 45.295 (rad/m). The black arrows represent the plane wave incident with a Gaussian beam, and the red arrows denote transmitted beams. It can be seen that, apart from the acoustic waves propagating along red arrows, there are also a few parts of energy traveling along other directions, which has been interpreted in detail in reference [27].